\documentclass{osa-article}
\journal{osajournal}
\articletype{Research Article}
\usepackage{amsmath,graphicx}
\usepackage{dcolumn}
\usepackage{bm}
\usepackage{xcolor}
\usepackage[style=nature, backend=biber]{biblatex}
\usepackage{hyperref}

\hypersetup{citecolor=blue,colorlinks=true,linkcolor=blue, linktocpage}

\begin{document}

\title{Phase-locking of lasers with Gaussian coupling}

\author{Andra Naresh Kumar Reddy,\authormark{1,$^\dagger$} Simon Mahler,\authormark{1,$^\dagger$} Alon Goldring,\authormark{1} Vishwa Pal,\authormark{2} Asher A. Friesem,\authormark{1} and Nir Davidson\authormark{1,*}}

\address{\authormark{1}Department of Physics of Complex Systems, Weizmann Institute of Science, Rehovot 7610001, Israel\\
\authormark{2}Department of Physics, Indian Institute of Technology Ropar, Rupnagar 140001, Punjab, India\\
\authormark{$^\dagger$}These authors contributed equally to this work.}

\email{\authormark{*}nir.davidson@weizmann.ac.il} 



\section*{Abstract}
A unique approach for steady in-phase locking of lasers in an array, regardless of the array geometry, position, orientation, period or size, is presented. The approach relies on the insertion of an intra-cavity Gaussian aperture in the far-field plane of the laser array. Steady in-phase locking of $90$ lasers, whose far-field patterns are comprised of sharp spots with extremely high power density, was obtained for various array geometries, even in the presence of near-degenerate solutions, geometric frustration or superimposed independent longitudinal modes. The internal phase structures of the lasers can also be suppressed so as to obtain pure Gaussian mode laser outputs with uniform phase and overall high beam quality. The approach could potentially improve the performances of recently developed laser simulators that are used for solving various computational problems.

\section*{Introduction}
Phase-locking of laser arrays, where all the lasers have the same frequency and same constant relative phase, plays an important role in many applications such as obtaining high laser output power with high beam quality~\autocite{Shalaby09,Mahler20,Roadmap21}, and beam shaping of lasers~\autocite{Shalaby09,Tradonsky21,Tradonsky17,Roadmap21}. Phase-locking of lasers also serves as a powerful tool to simulate and investigate a variety of phenomena including computational and optimization problems~\autocite{Mahler20,Marandi14,Berloff17,Takeda17}, spin systems~\autocite{Nixon13,Berloff17,Pal20,Miri20} and more. 

Many coupling techniques have been developed during the past decades for efficiently phase-locking lasers in an array. These include coupling with Fourier and Talbot diffraction~\autocite{Glova03,Tradonsky17}, one-dimensional coupling with spherical or cylindrical lenses~\autocite{Mahler20}, coupling by manipulating mirrors, and coupling with diffractive elements~\autocite{Zhou04,Mahler20,Cheung08}. With all these, the coupling strongly depends on the array geometry, the size of the lasers, and the distance between them~\autocite{Mahler19}, all of which must be accurately controlled and kept strictly constant in order to converge to the minimal-loss phase-locking state~\autocite{Mahler20,Pal20}. Accordingly, disordered arrays of lasers cannot be phase-locked in the in-phase state with these coupling techniques. Other complicated array geometries such as Kagome and triangular will inherently suffer from inconsistencies between nearest and next-nearest neighbors coupling phase~\autocite{Pal20,Nixon13,Chalker92}. Even a simple square array geometry requires precise setting and positioning of the coupling elements that could be optimized only for a specific separation between the lasers~\autocite{Mahler20,Tradonsky17}. 

Here, we present a different and relatively simple coupling technique based on a Gaussian coupling function. The positive Gaussian coupling function ensures that all the lasers are always positively coupled regardless of the array geometry, position, orientation, period or size, thereby removing frustration associated with other coupling techniques where the sign of the coupling changes with the distance and with the array geometry. About $90$ independent lasers formed in a degenerate cavity laser were efficiently in-phase locked by a Gaussian aperture that is inserted in the far-field plane. Steady in-phase locking of lasers  was also obtained for the Kagome and random arrays of lasers, even in the presence of near-degenerate solutions, frustration or superimposed longitudinal modes. Moreover, internal phase structures of the lasers were suppressed, so as to result in pure Gaussian laser outputs with uniform phase and improved overall beam quality. 

\textbf{Basic principle.} Far-field coupling of lasers relies on a simple principle: an optical element (aperture) placed in the (Fourier) far-field plane of an array of $N$ lasers will result in the convolution of the array field with the coupling function of the aperture (i.e. the Fourier transform of the aperture transmission function $T(x,y)$~\autocite{Tradonsky17}). The resulting convolved field corresponds to the coupling matrix of the lasers. The $N$ eigenmodes and eigenvalues of the coupling matrix represent the different possible phase-locking states with corresponding loss. The minimal loss phase-locking state is defined as the eigenmode that has the lowest loss (i.e. 1-eigenvalue). 

To calculate the coupling between a selected laser and the other lasers, we convolved the intensity distribution of the selected laser with the Fourier transform of the aperture transmission function. Details about the calculation are given in the Appendix~\ref{sec:AppA}. Figure~\ref{fig:1_coupling_functions} shows the coupling functions between a selected laser and the others in a square array of period $a$, for a binary aperture (red dashed curve) and a Gaussian aperture (black solid curve), where a Gaussian aperture is an aperture whose transmission function is a Gaussian function.
 
\begin{figure}[!ht]
\centering\includegraphics[width=0.55\textwidth]{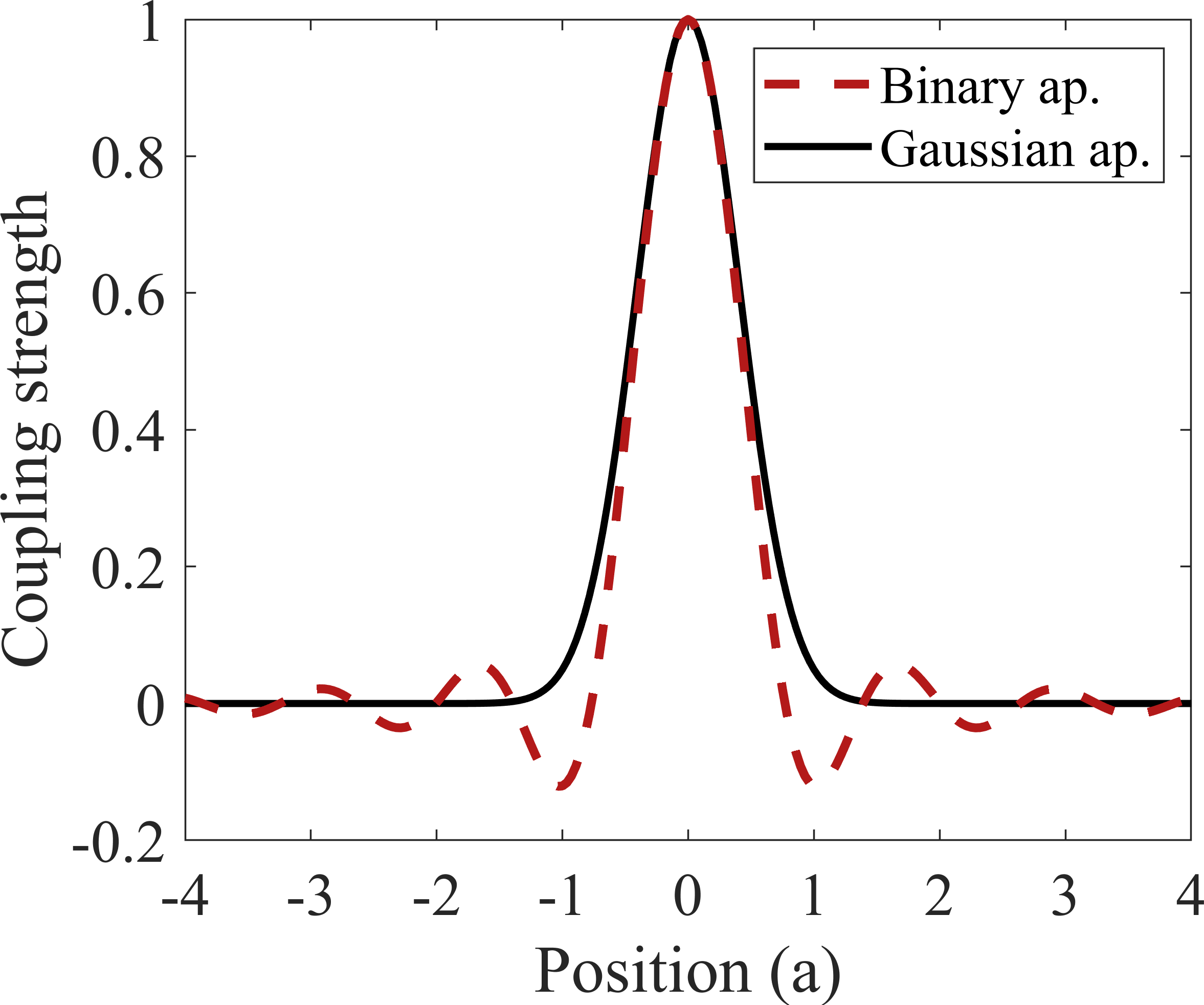}
\caption{\textbf{Binary and Gaussian apertures coupling functions.} With a binary aperture, the sign, range, and strength of the coupling depends on the period, orientation and size of the lasers. With a Gaussian aperture, the coupling is always positive.}
\label{fig:1_coupling_functions}
\end{figure}

As evident, the coupling function of the binary aperture exhibits a periodic $sinc(x)=sin(x)/x$ function with positive and negative values. The period and size of the $sinc$ depends on the parameters of the array of lasers such as the position, orientation, period or size of the lasers, and on the size of the aperture, see Appendix~\ref{sec:AppA}, causing positive and negative coupling between lasers at different distances. This gives rise to frustration in certain array geometries (e.g. triangular or Kagome arrays)~\autocite{Pal20}. Similar results would be obtained with other coupling techniques such as Talbot diffraction, saturable absorber and phase elements.

The coupling function of the Gaussian aperture is always a positive Gaussian function. Thereby, with the Gaussian aperture, lasers are only positively coupled. 
A continuous control of the range of the positive coupling, from no coupling, to nearest neighbors, to next nearest neighbors, to all the way to mean field (all to all) coupling can be implemented by simply varying the size of the Gaussian aperture, see  Appendix~\ref{sec:AppA}. Specifically, by varying the size of the Gaussian aperture, it is possible to control the coupling strength and range while keeping the coupling real and positive for any direction and distance, ensuring in-phase locking. 

Gaussian coupling decays faster with distance than binary coupling, so undesired next nearest neighbor coupling can be suppressed compared to nearest-neighbor coupling, unlike the power-law decay of the coupling strength in the other techniques that often leads to frustration~\autocite{Pal20,Nixon13}. The Gaussian coupling can thus be exploited for simulating spins with positive coupling as a unique, clean physical simulator with no degeneracy in the ground state. In the following, we describe and demonstrate our experimental arrangement and the versatility and robustness of the Gaussian coupling with a variety of laser arrays of different geometries, all of which are stably phase locked in the in-phase state. 

\section*{Results}
\begin{figure}[h!]
\centering\includegraphics[width=0.75\textwidth]{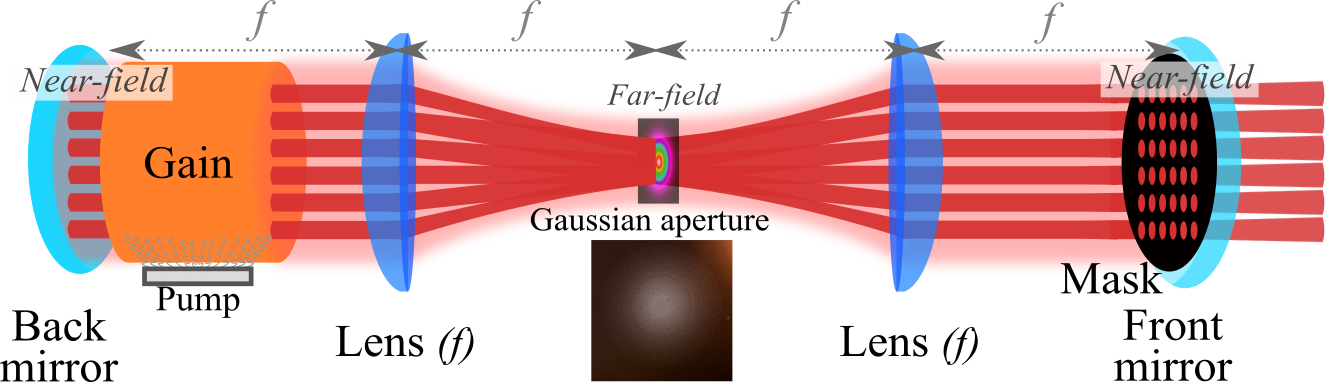}
\caption{\textbf{Experimental arrangement for forming a Gaussian coupled array of lasers.} A Gaussian aperture is inserted at the far-field (Fourier) plane of a degenerate cavity laser and a mask of holes is inserted at the near-field plane.}
\label{fig:2_exp_arrangement}
\end{figure}

\textbf{Experimental arrangement and results.} Our experimental arrangement for forming and coupling an array of lasers is schematically presented in Fig.~\ref{fig:2_exp_arrangement}. It is comprised of a self-imaging degenerate cavity laser (DCL)~\autocite{Arnaud69}, with a mask of holes for forming independent lasers in an array~\autocite{Mahler19}, and a Gaussian aperture to couple them.  

The DCL is comprised of a back flat mirror of high reflectivity. Next to it, a Nd:YAG crystal rod, lasing at $\lambda=1064$ nm, serves as a gain medium that is pumped by a flashlamp. At focal distance $f$ from the back mirror, a spherical lens of focal length $f=40$ cm is inserted. A second identical lens is inserted at distance $2f$ away from the first lens. The two lenses form a $4f$ telescope configuration. Finally, at focal distance $f$ from the second lens, a front mirror of reflectivity $R=80\%$, acts as an output coupler. Additional details about the experimental arrangement are given in the Methods and in Appendix~\ref{sec:AppB}.

The $4f$ telescope ensures that any field distribution at the back mirror is imaged onto the front mirror. These fields are denoted near-field. The field midway between the lenses is denoted far-field and is equivalent to the Fourier transform of the near-field~\autocite{vonBieren71}. Both near-field and far-field planes are physically accessible. A mask of holes with desired shape, geometry and sizes can be inserted at the near-field plane to generate independent lasers (with Gaussian modes)~\autocite{Nixon13,Tradonsky17}. In all our experiments, the diameter of the lasers was $d=200$ $\mu$m, and the period of the array was $a=300$ $\mu$m. A binary or Gaussian aperture can be inserted at the far-field (Fourier) plane to couple them~\autocite{Mahler19,Tradonsky17}. 

An external imaging system (shown in Appendix~\ref{sec:AppB}), imaged both the near-field and far-field planes onto a camera. Several Gaussian apertures of different diameters were manufactured by a laser writing process with a resolution of $5\pm0.5$ $\mu$m, see  Appendix~\ref{sec:AppD}. For a Gaussian aperture, the diameter is defined as the full width at $1/e^{2}$ of the Gaussian transmission function, whereas for a binary aperture, the diameter is defined as the diameter where the transmission is one.   

\begin{figure}[h!]
\centering\includegraphics[width=0.55\textwidth]{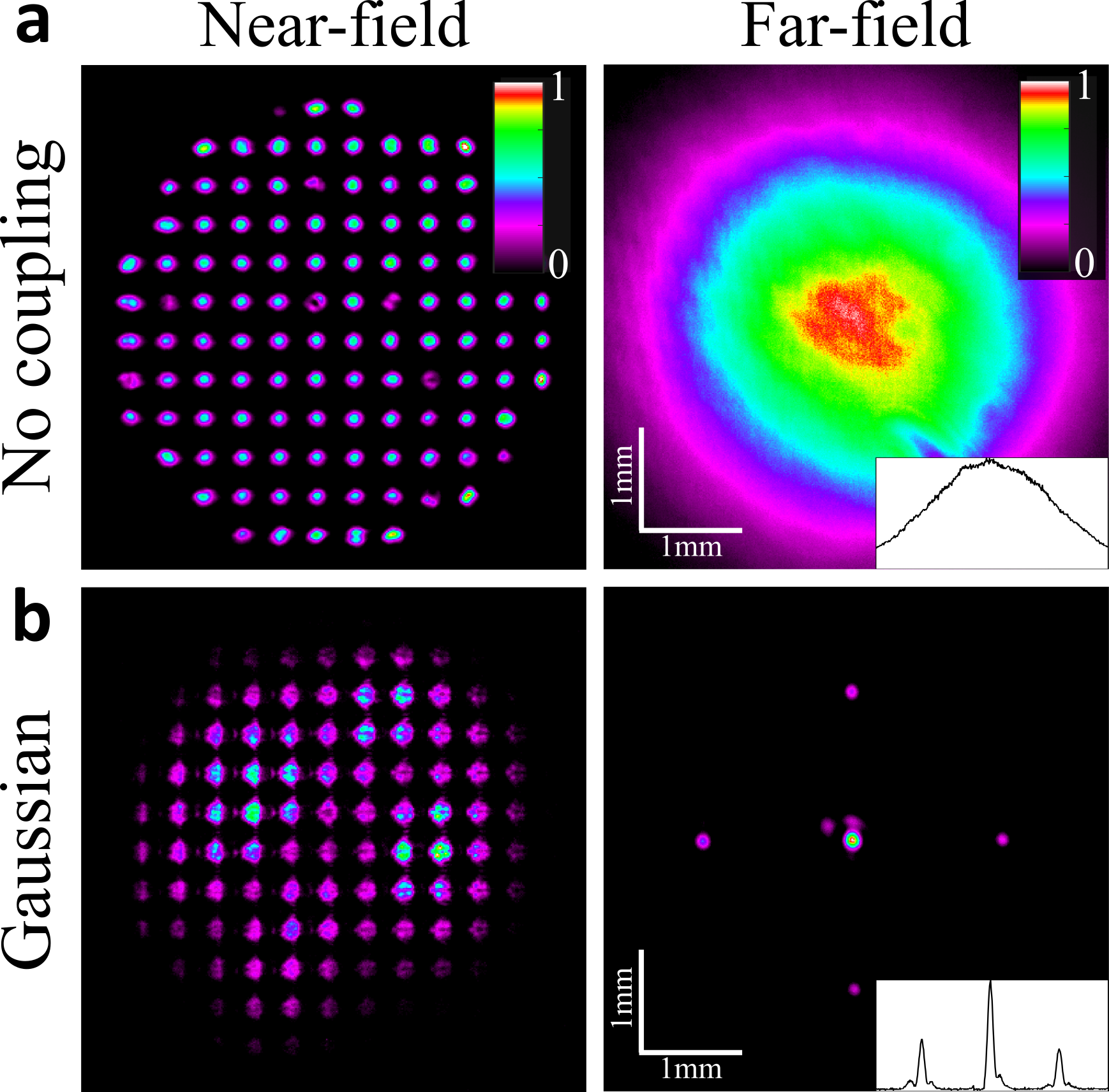}
\caption{\textbf{Gaussian coupling of lasers in a square array.} Detected near-field and far-field intensity distributions. \textbf{a} Without aperture, and \textbf{b} with a Gaussian aperture of diameter $D=1$ mm. As evident, with the Gaussian aperture, the lasers are strongly coupled and all have the same phase (in-phase locking). Insets - Horizontal cross section at the far-field center.}
\label{fig:3_G_coupling}
\end{figure}

\begin{figure}[h!]
\centering\includegraphics[width=0.95\textwidth]{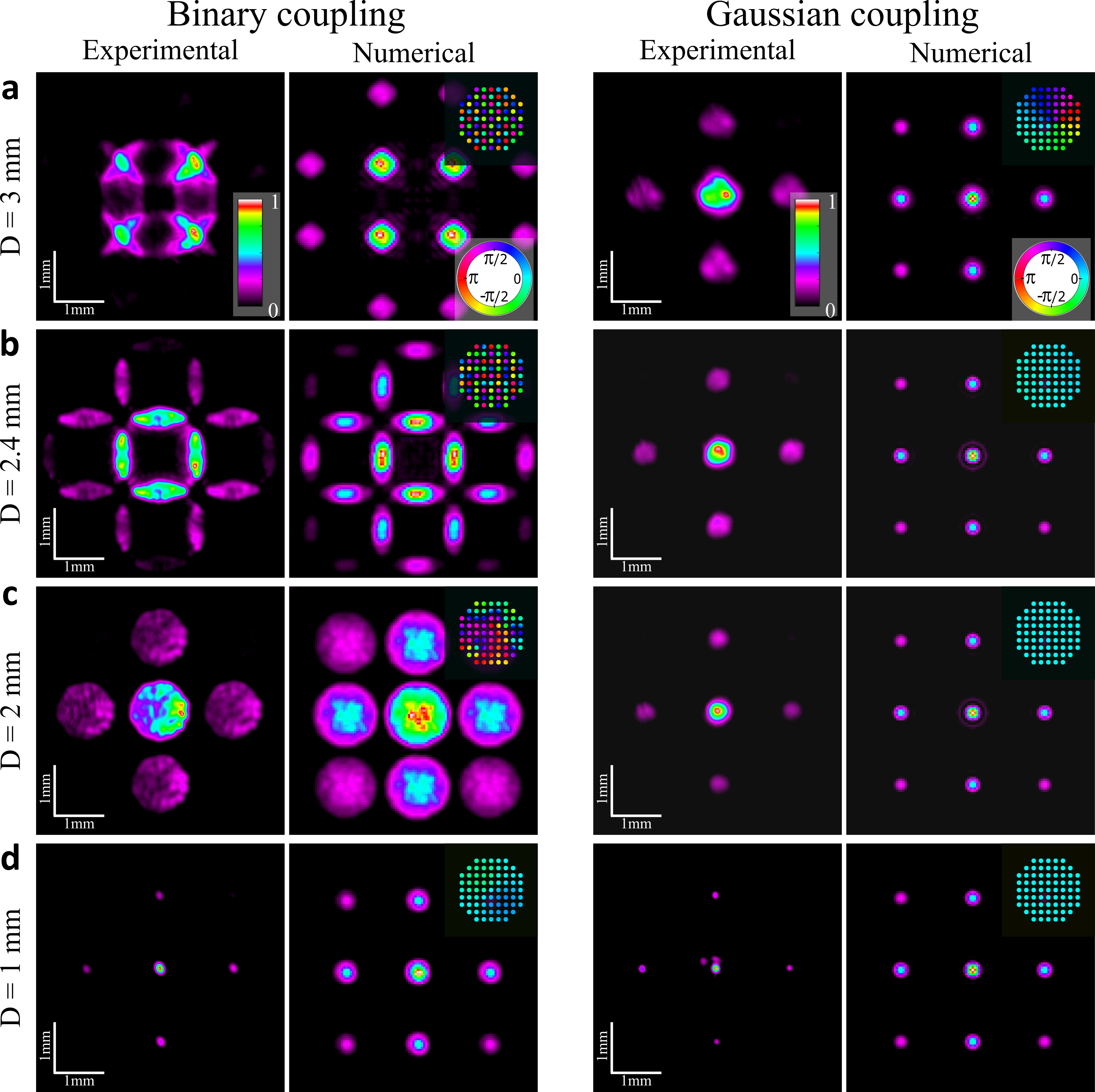}
\caption{\textbf{Binary versus Gaussian couplings.} Detected and numerically simulated far-field intensity distributions of the lasers in a square array for different diameters $D$ of intra-cavity binary apertures (left columns) and Gaussian apertures (right columns). \textbf{a} $D=3$ mm, \textbf{b} $D=2.4$ mm, \textbf{c} $D=2$ mm, \textbf{d} $D=1$ mm, Insets - Simulated phases distribution of the lasers. As evident, with binary apertures, the lasers phase lock with either positive or negative or other coupling, depending on the aperture size. With Gaussian apertures, the lasers always lock in-phase, regardless of the size of the aperture.}
\label{fig:4_B_vs_G_coupling}
\end{figure}

\begin{figure}[h!]
\centering\includegraphics[width=0.55\textwidth]{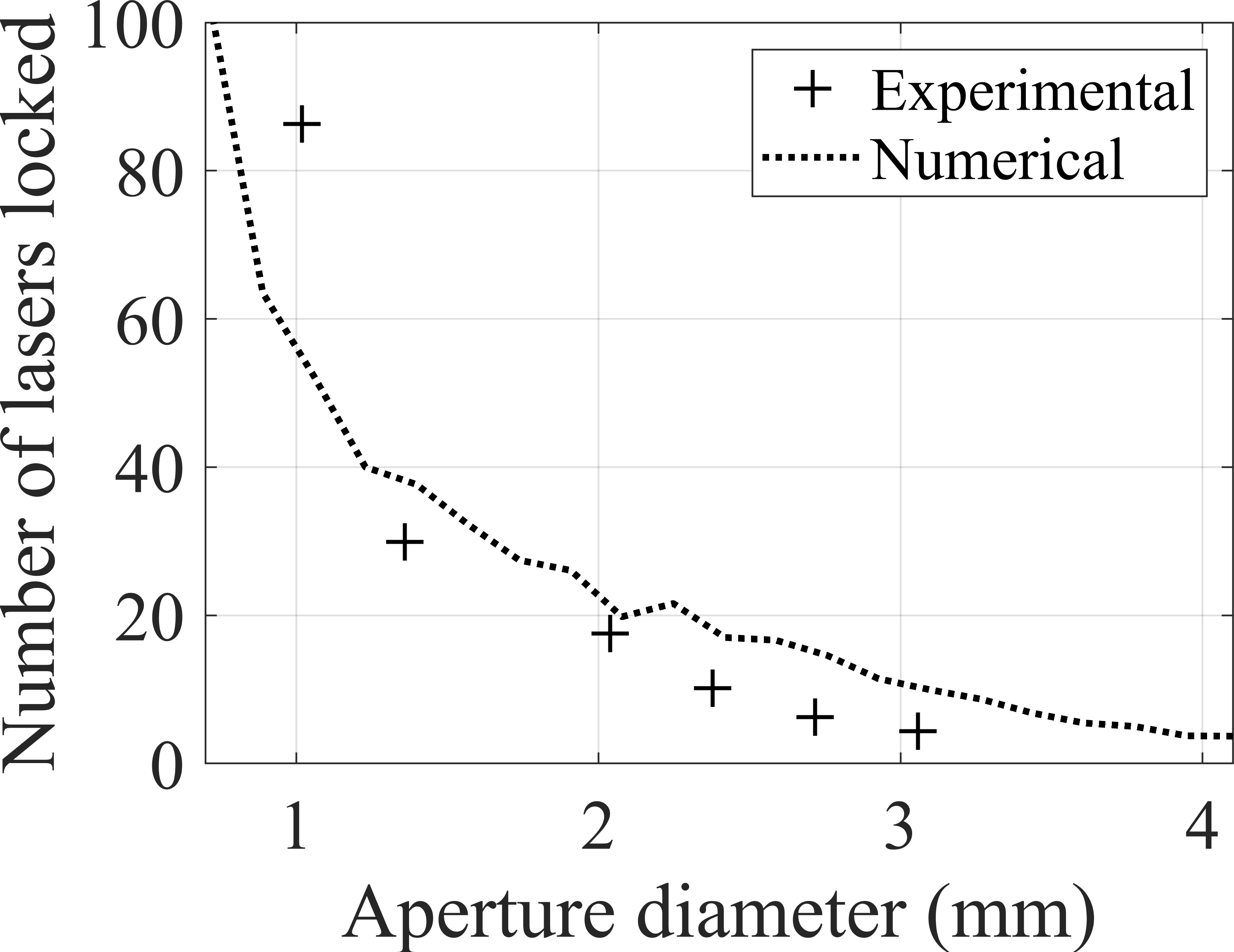}
\caption{\textbf{The number of lasers that are in-phase locked in a square array as a function of Gaussian aperture diameter.} As evident, the number of lasers monotonically increases as the diameter of the aperture decreases.}
\label{fig:5_N_vs_D_G_ap}
\end{figure}

\begin{figure}[h!]
\centering\includegraphics[width=0.55\textwidth]{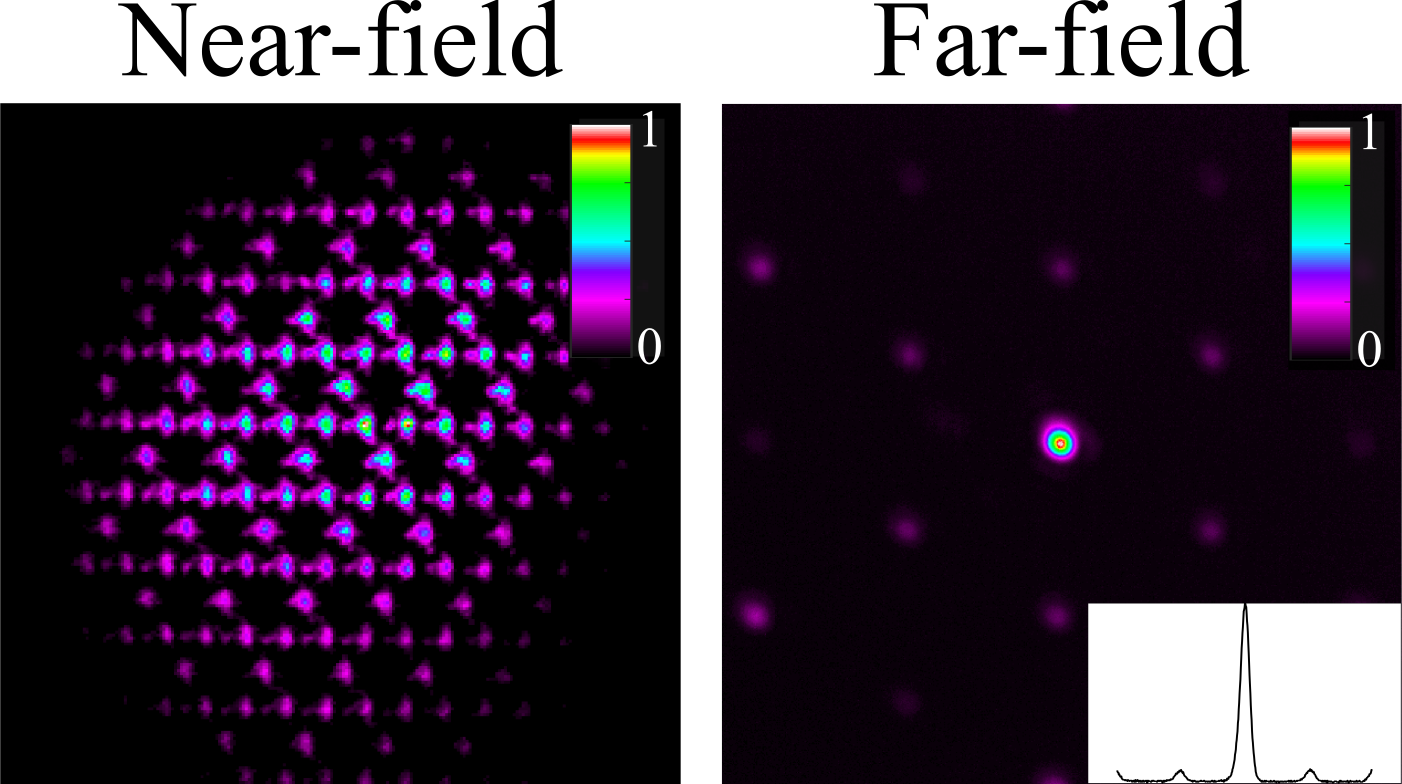}
\caption{\textbf{Gaussian coupling in Kagome array of lasers.} Detected near-field and far-field intensity distributions of in-phase locked lasers, for a Kagome array of lasers using a Gaussian aperture. Inset - Vertical cross section at the far-field center.}
\label{fig:6_Kagome_phs_lck}
\end{figure}

We performed a series of experiments to demonstrate the efficacy of Gaussian coupling in square (Figs. \ref{fig:3_G_coupling}, \ref{fig:4_B_vs_G_coupling} and \ref{fig:5_N_vs_D_G_ap}), Kagome (Fig.~\ref{fig:6_Kagome_phs_lck}) and quasi-random array of lasers ( Appendix~\ref{sec:AppC}). We also performed numerical simulations by using a special algorithm~\autocite{Mahler20_2} that combines the Fox-Li and the Gerchberg-Saxton algorithms, where a saturable gain function and detuned random phase field mimicked the path of the light inside our experimental arrangement, see Methods.

Figure~\ref{fig:3_G_coupling} shows the detected near-field and far-field intensity distributions for a square array without an intra-cavity aperture, Fig.\ref{fig:3_G_coupling}a, and with a Gaussian aperture, \ref{fig:3_G_coupling}b, of $D=1$ mm  that provides  Gaussian coupling.

As evident, without an intra-cavity aperture, a broad far-field intensity distribution is observed where the lasers are uncoupled with different independent phases ranging from $[-\pi$ to $\pi]$. With the Gaussian aperture, the far-field intensity distribution is composed of a bright high-energy sharp spot at the center (zero-order spot), surrounded by weak spots. This indicates that the lasers are positively coupled and are locked in-phase, where all the lasers have the same constant phase. The four weak spots correspond to the first-order spots and are equally separated by a distance $L_{1,0}=\lambda f/a=1.4$ mm ($\lambda$ the wavelength, $f$ the focal length of the lenses and $a$ the period of the array) from the zero-order central spot~\autocite{Mahler20_2}. 

Results showing the differences between binary and Gaussian couplings are presented in Fig.~\ref{fig:4_B_vs_G_coupling}. They show the detected and numerically simulated far-field intensity distributions for a square array of lasers with binary and Gaussian apertures of different diameters $D=[3, 2.4, 2,$ and $1]$ mm.

The far-field intensity distributions for the binary coupling (left columns) correspond to different phase-locking states that vary as the diameter of the intra-cavity binary aperture D is reduced: out-of-phase locking for D=3 mm, mixed phase locking for D=2.4 mm, very short range in-phase locking for D=2 mm, and eventually long-range in-phase locking for D=1 mm (clearly seen also in the simulated near-field phases shown in the insets). In contrast, for the Gaussian coupling (right columns) robust in-phase locking is obtained for all diameters of the Gaussian intra-cavity Gaussian aperture, as clearly seen from the far-field intensity distributions and  the simulated near-field phases, shown in the insets.   

These results indicate that in order to obtain long range in-phase locking, the binary aperture diameter needs to be precisely tuned to the array geometry while the Gaussian aperture diameters can range over many sizes. Frustrated phase-locking or very short range locking can also arise with binary apertures but not with Gaussian apertures. Although our DCL has hundreds of independent longitudinal (temporal) modes~\autocite{Mahler20}, the Gaussian aperture provides in-phase locking of all mode, avoiding superposition of phase-locking states~\autocite{Mahler20}.

The range of phase-locking of lasers can be quantified as the ratio of the distance between nearby far-field diffraction spots, to their width ~\autocite{Mahler20_2}. For a two-dimensional square array,  the number of mutually phase locked lasers, $N$, is the square of this range. Fig.~\ref{fig:5_N_vs_D_G_ap} presents $N$ deduced from the experimental and the numerical far-field intensity distributions in Fig.~\ref{fig:4_B_vs_G_coupling}, for different Gaussian aperture diameters. As evident, $N$ increases monotonically as the aperture diameter decreases and reaches about $90$ for $D=1$ mm.

Efficient in-phase locking of lasers in a Kagome array is a non-trivial task. The Talbot coupling method suffers from significant loss and the binary coupling method requires a precisely tuned and extremely small aperture to avoid superposition of degenerate lowest loss states. For example, with negative coupling, lasers exhibit frustrations due to the $\propto2^{N}$ degenerate low-loss phase-locking states~\autocite{Pal20,Nixon13,Mahler20_2, Chalker92}. Figure~\ref{fig:6_Kagome_phs_lck} shows the near-field and far-field intensity distributions of a Kagome laser array, with Gaussian coupling. The sharp central peak in the far-field (see inset) verifies stable in-phase locking of most lasers in the array. In-phase locking with Gaussian coupling was also successfully demonstrated in a quasi-random array of lasers, see Appendix~\ref{sec:AppD}. These results clearly demonstrate superiority over other coupling techniques for phase-locking complex laser array geometries. 

In addition to providing coupling between different lasers in the array, the far-field Gaussian aperture also acts as a spatial filter, ensuring that each laser in the array is a pure Gaussian mode with no internal structure, see Appendix~\ref{sec:AppF}. Pure single-mode Gaussian lasers are advantageous for many phase-locking applications and in particular in spin simulators and solvers~\autocite{Pal20}. 

Finally, we studied the loss induced by the intra-cavity far-field Gaussian aperture. For a $1$ mm diameter aperture, capable of phase-locking all $90$ lasers in the array, we obtained a round-trip loss of about $90\%$, dominated by the high-order diffraction lobes that are completely blocked by the aperture. We obtained a much smaller round-trip loss of about $35\%$, by resorting to a combined aperture containing $5$ Gaussian apertures one centered at the zero-order lobe and four other at the first-order diffraction lobes. The need to center each Gaussian aperture to its diffraction lobe does reduce the tolerance of this more efficient configuration to the array geometry. For more details, see Appendix~\ref{sec:AppE}.

\section*{Conclusion}
To conclude, we presented a simple, robust and efficient approach for steady in-phase locking of lasers by means of Gaussian coupling. We showed that the Gaussian coupling function is always positive as opposed to other coupling functions, where frustration and phase oscillations in the lasers can arise due to the changes of sign in the coupling function. The range of the coupling as well as the number of lasers that are phase locked were gradually controlled from no coupling to mean-field (all to all) coupling by changing the diameter of the Gaussian aperture. Steady in-phase locking of lasers was also demonstrated in a Kagome and quasi-random laser arrays. Such in-phase locking can be achieved within less than $100$ ns and with an high power density sharp spot in the far-field \cite{Mahler20}.   

In recent years, there is a growing interest on exploiting laser systems for solving computational problems, such as phase retrieval problems~\autocite{Tradonsky19,Zhang16}, $XY$ spin systems~\autocite{Nixon13,Pal20,Mahler20,Berloff17,Takeda17} and coherent Ising machine~\autocite{Pierangeli19,Yamamoto17,Marandi14}. Our findings and results should be of interest for elucidating the physical mechanism of solving such computational problem with lasers.  

Finally, we note that our approach is not limited to Gaussian coupling function by the use of Gaussian aperture but can be applied to any desired coupling function (as long as it depends only on the relative displacement of the lasers) using its Fourier relation to the aperture transmission function $T(x,y)$~\autocite{Tradonsky17}) . Useful examples include rectangular coupling (via a $sinc(x)=sin(x)/x$ aperture), Bessel coupling (via a ring aperture) or double exponential coupling (via a Lorentzian aperture). 

\begin{backmatter}
\section*{Methods}
\textbf{Experimental arrangement and technical details.} In this section, we elaborate on our experimental arrangement and external imaging system, shown in Appendix~\ref{sec:AppB}. As shown, our DCL is comprised of a back flat mirror of reflectivity $R=99.5\%$ at $1064$ nm. Next to it, a Nd:YAG crystal rod of $10.9$ cm length and $1$ cm diameter served as a gain medium, lasing at $\lambda=1064$ nm. The gain medium was optically pumped by a quasi-CW $100$ $\mu$s pulsed flash-lamp operating at $1$ Hz repetition rate to avoid thermal effects. At focal distance $f$ from the back mirror, a spherical lens of $5.08$ cm diameter and focal length $f=40$ cm was inserted. A second identical lens was inserted at focal distance $2f$ away from the first lens, so the two lenses formed a $4f$ telescope configuration. Finally, at focal distance $f$ from the second lens, a front mirror of reflectivity $R_{o.c}=80\%$, acted as an output coupler. 

The $4f$ telescope ensured than any field distribution at the back mirror is precisely imaged onto the front mirror. The fields at the back and front mirrors are thereby equivalent and denoted as near-field. The field midway between the lenses is denoted far-field and is equivalent to the Fourier transform of the near-field. A mask of holes of period $a=300$ $\mu$m can be inserted in the near-field plane and a Gaussian (or binary) aperture in the far-field plane.

An external imaging system imaged both the near-field and far-field planes onto a camera. For that, a lens of focal length $f_{2}=20$ cm was inserted at focal distance $f_{2}$ from the front mirror. A beam splitter splits the laser light into two arms. In the first arm, the laser light propagates in free space. In the second orthogonal arm, a lens of focal length $f_{3}=10$ cm is inserted at a focal distance $f_{2}+f_{3}$ from the preceding lens. Then, a second beam splitter recombines the light from the two arms onto a camera, placed at focal distance from the lenses. The first (horizontal) arm imaged the far-field plane of the DCL, whereas the second (orthogonal) arm imaged the near-field plane of the DCL. To avoid spatial overlap of laser light from the two arms on the camera, a transverse shift was introduced by slightly rotating the second beam splitter. 

\textbf{Numerical simulation.} The numerical simulations of the manuscript (Figs.~3, 4 5 and 7) were performed with an iterative algorithm that combines the Fox-Li and the Gerchberg-Saxton algorithms, where a saturable gain function simulated the lasing intensity dynamics. The iterative algorithm numerically mimics the free space propagation of the laser field inside the DCL. One iteration in the algorithm corresponds to one round-trip of the laser field in the DCL. 

For the numerical simulations, we start with a matrix $E_{ij}$, $i,j\in [1....L^{2}]$ with $L^{2}$ the number of pixels in the laser field to represents the spatial distribution of the laser field with uniform intensities and random phases. Then, a mask of holes $Mask_{ij}$  is formed with transmission $T_{i_{out}j_{out}}=0$ everywhere except inside a hole where the transmission is $T_{i_{in}j_{in}}=1$. This $Mask_{ij}$ corresponds to the near-field mask of holes inserted in the DCL for forming an array of independent lasers. It can be a square, Kagome or random arrays of holes of period $a$. An aperture $Aperture_{ij}$ is then used with transmission $0$ everywhere except inside the aperture. This $Aperture_{ij}$ corresponds to the far-field aperture of diameter $D$ inserted in the DCL for coupling the lasers. It can be either a binary aperture with a binary transmission function of $1$ or a Gaussian aperture with a Gaussian transmission function. Finally, a saturable gain function $G_{ij}=\frac{G_{0}}{1+|E_{ij}|^{2}/I_{sat}}$ that mimics the gain and loss inside the DCL is also used where $G_{0}$ is the constant pump value and $I_{sat}$ the saturation intensity.

During each iteration, the laser field $E_{ij}$ is multiplied by the saturable gain function $G_{ij}$, then it is Fourier transformed (far-field $FF_{ij}$), then is multiplied by the aperture $Aperture_{ij}$, then is inversely Fourier transformed and multiplied by the mask of holes $Mask_{ij}$ (near-field $NF_{ij}$). The resulting laser field is used for the next iteration as $E_{ij}=NF_{ij}$. This process is repeated $T$ times, where $T$ is the number of iterations.

Initially, the laser field has uniform intensities with $|E_{ij}|^{2}<<I_{sat}$, so the gain inside the DCL is uniform with value $\approx G_{0}$. After one round-trip, most of the pixels $i_{out}j_{out}$ in the laser field have zero transmission in the near-field mask of holes $Mask_{ij}$ so their intensity is  $|E_{i_{out}j_{out}}|^{2}=0$. Only the group of pixels that is inside a hole, corresponding to a laser in the array, have non zero intensities $|E_{i_{in}j_{in}}|^{2}>0$. Thereby, the pixels $i_{in}j_{in}$ have a  much higher gain $G_{i_{in}j_{in}}$ during the first iterations (i.e. during the first round-trips in the DCL). When the loss inside the DCL equals the gain, lasing occurs with gain and power saturation. Similarly, when $|E_{i_{in}j_{in}}|^{2}=I_{sat}$, lasing occurs in the numerical simulation and the gain saturates. Thereby, the iterative algorithm accurately simulates the near-field and far-field distributions of the DCL as well as the lasing intensity and loss distributions. 

In most of  numerical simulations, the number of iterations was $T=115$, the pixel size was $dr=20$ $\mu$m, the wavelength was $1.064$ $\mu$m, the focal length of the lenses was $40$ cm, $G_{0}=15$, $I_{sat}=1000$, and $L^{2}=500$ such that the field size was a $1\times1$ cm square. To mimic the different longitudinal modes present in the DCL, the simulations were averaged over $50$ realizations for the near-field and far-field intensity distributions in Figs. 4 and 5, and for the power efficiency results in Figs. 12 and 13, and over $1$ realization for the near-field phase distribution in Figs. 3 and 7.

\section*{Acknowledgments} The authors thank Chene Tradonsky and Sagie Gadasi for valuable help and thank the Israel Science Foundation (501100003977) for their support.
\end{backmatter}

\printbibliography
\newpage 

\begin{backmatter}
\renewcommand{\thesection}{\Alph{section}}   
\section{Appendix A: Calculation of the coupling function}
\label{sec:AppA}
In Fig.~\ref{fig:1_coupling_functions} of the manuscript we briefly presented basic coupling functions of the binary and Gaussian apertures. In this section, we describe the calculation process for determining the coupling functions, and show coupling results for different binary and Gaussian aperture diameters.

The coupling function was calculated by using a simple algorithm that mimics the propagation inside our degenerate cavity laser (DCL) of a single Gaussian mode laser in an array of lasers. First, an array of $N$ pure Gaussian mode lasers is initialized as a matrix $Mask$ (intensity distributions in left insets in Fig.~\ref{fig:S1_Coupling_functions}). 

\begin{figure}[!ht]
\centering\includegraphics[width=0.85\textwidth]{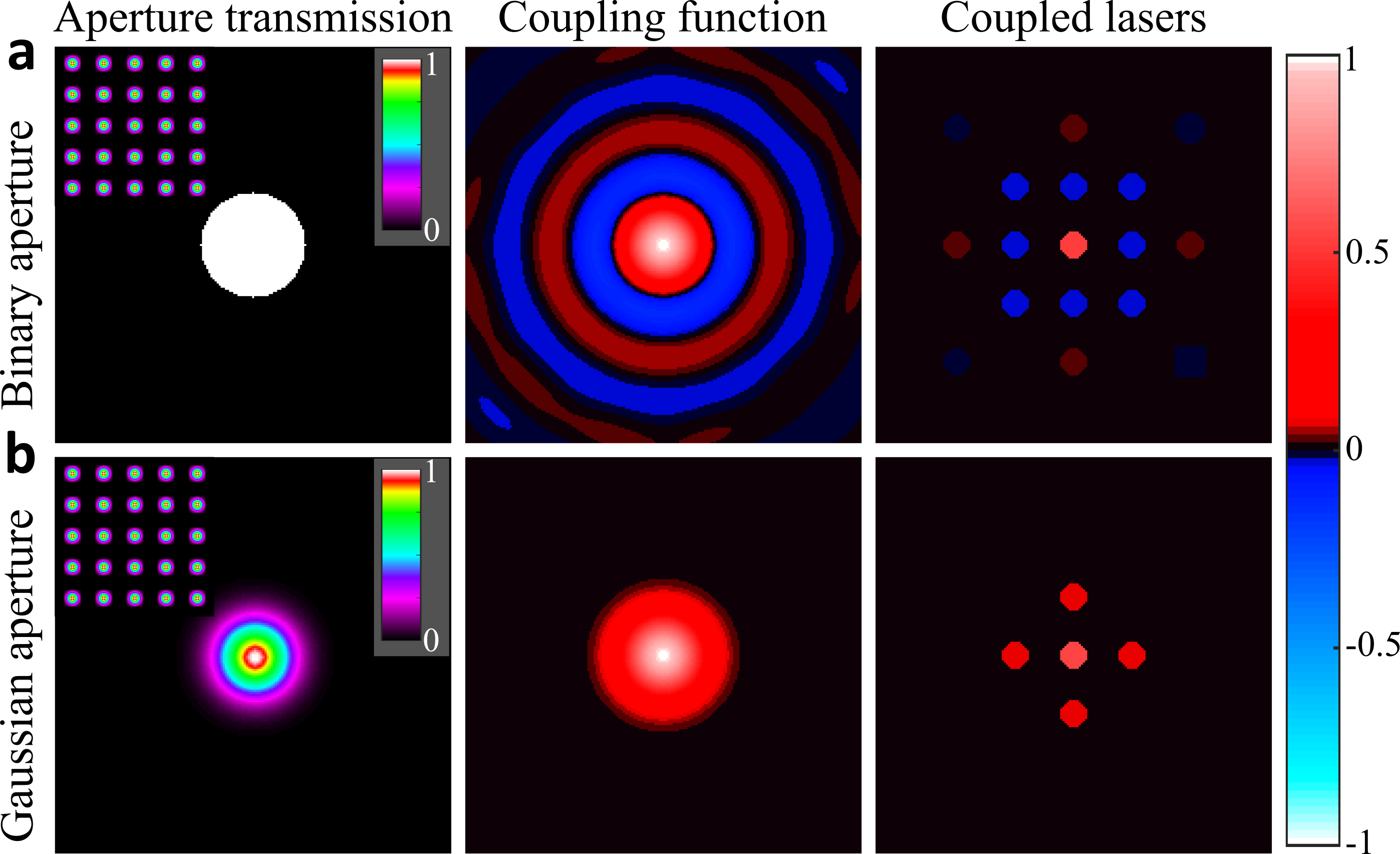}
\caption{\textbf{Coupling functions of binary and Gaussian apertures.} \textbf{a} Binary coupling functions. \textbf{b} Gaussian coupling function. Left - Transmission distributions of the intra-cavity far-field aperture that couples lasers. Inset - Near-field intensity distributions of the lasers in a square array where the selected laser is the one at the center. Middle - Coupling function of the selected laser after a cavity round-trip. Right - Coupling strength distribution between the selected laser and the others.}
\label{fig:S1_Coupling_functions}
\end{figure}

Then a selected laser, preferably in the center of the array, is kept and all the other lasers are removed. The resulting matrix of the selected laser, denoted as $Laser$, is then Fourier transformed, multiplied by the far-field coupling $Aperture$ (either binary or Gaussian or none), and is inversely Fourier transformed. The resulting field, denoted as $Coupler$, corresponds to the coupling function of the selected $Laser$, as shown in the middle column of Fig.~\ref{fig:S1_Coupling_functions}. To obtain the coupling strength distribution between the selected laser and the others, the coupling function $Coupler$ was multiplied by the array of lasers $Mask$, where the coupling distribution within each laser was averaged to remove internal variations. This calculation simulates a single round-trip propagation inside the DCL without gain. A more elaborate and accurate algorithm, with gain effects, that exactly mimics the lasing output of the DCL was used to perform numerical simulations, see Methods.

Figure~\ref{fig:S1_K_fcs_vs_diam} shows the coupling function of the selected laser (first and third rows) and its coupling with the other lasers (second and fourth rows) for a square array of lasers and binary and Gaussian apertures of different sizes. As evident, the coupling function of the binary apertures exhibit a periodic $sinc(x)$ function with positive and negative values, except for the case where the size of the aperture is extremely small. The coupling function of the Gaussian aperture is always a positive Gaussian function.

\begin{figure}[!ht]
\centering\includegraphics[width=0.95\textwidth]{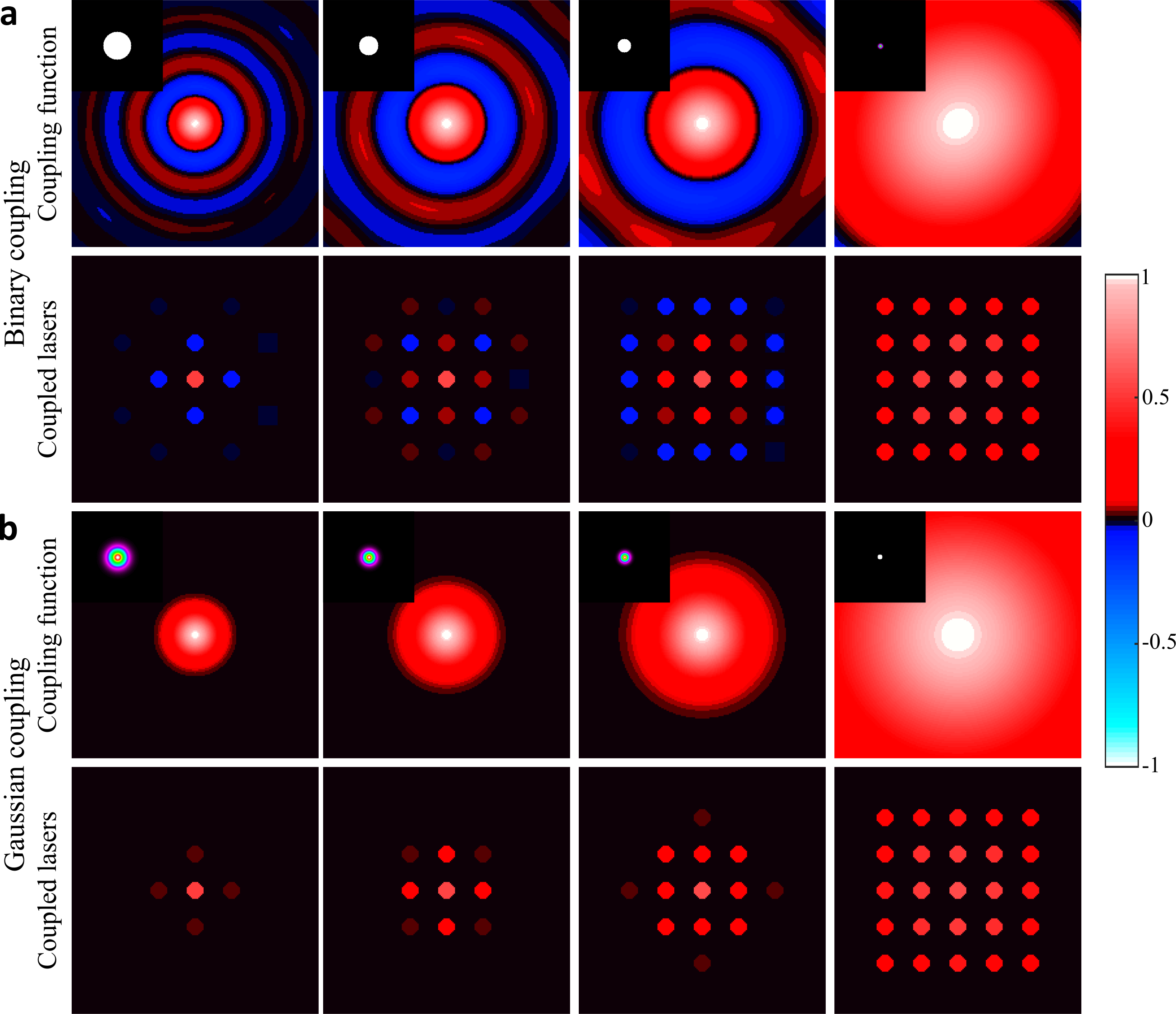}
\caption{\textbf{Binary and Gaussian coupling functions for different size of the apertures.} \textbf{a} Binary coupling. \textbf{b} Gaussian coupling. First and Third rows - Coupling function of the selected laser. Inset - Transmission distributions of a far-field aperture that couple lasers. Second and fourth rows - Coupling distribution between the selected laser and the others. As evident, with binary coupling, the sign, range, and strength of the coupling highly depends on the size of the far-field aperture and the lasers. With the Gaussian coupling, the coupling is always positive and with a shorter range.}
\label{fig:S1_K_fcs_vs_diam}
\end{figure}

The range of the coupling function is the inverse of the size of the aperture. As the aperture size decreases, the coupling range increases. Thereby, with a Gaussian aperture, by continuously varying the size of the aperture, one can continuously control the range of the coupling from no coupling, to nearest neighbors, to next nearest neighbors, to all the way to mean field (all to all) coupling. For the binary aperture, due to the positive and negative values of the coupling function, very specific and undesired types of coupling can be obtained. For example, in the third column, the selected laser is positively coupled with its nearest and next nearest neighbors but is negatively coupled with its next next nearest neighbors. Such couplings can lead to frustration, as in Fig.~\ref{fig:4_B_vs_G_coupling}b.

\newpage 
\section{Appendix B: Detailed experimental arrangement sketch}
\label{sec:AppB}
Figure~\ref{fig:S3_Exp_sketch_with_im_system} schematically shows the experimental arrangement and the imaging system used to perform all the experiments reported in the manuscript .

\begin{figure}[!ht]
\centering\includegraphics[width=0.95\textwidth]{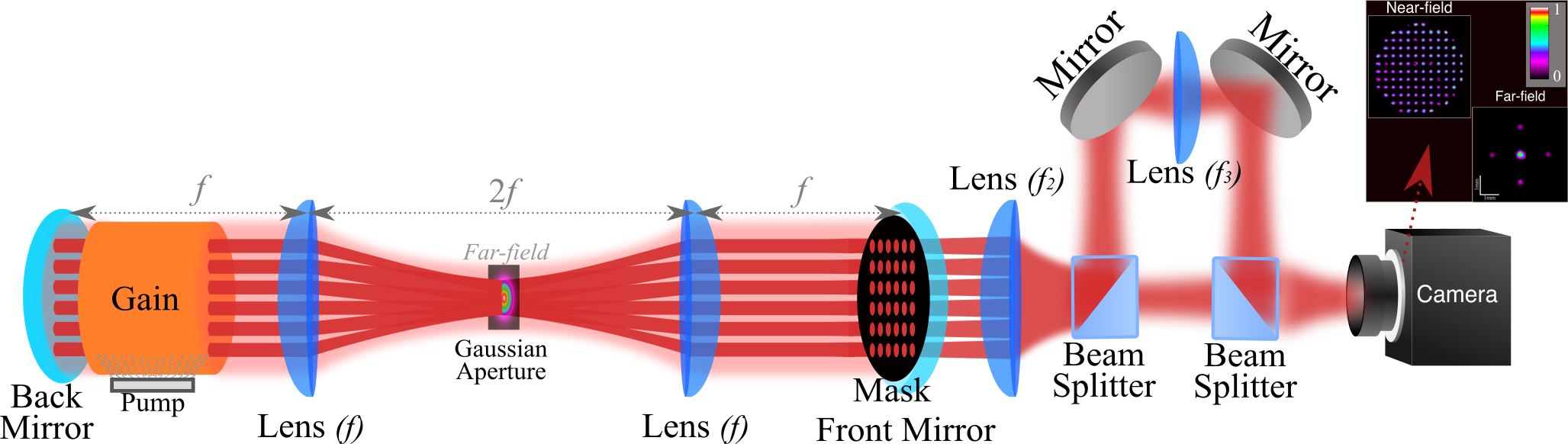}
\caption{Experimental arrangement of the degenerate cavity laser and the imaging system. A Gaussian aperture is inserted at the far-field (Fourier) plane of the laser cavity and a mask of holes is inserted at the near-field plane. An imaging system, outside the laser cavity, images both the near-field and far-field planes onto a camera.}
\label{fig:S3_Exp_sketch_with_im_system}
\end{figure}

\section{Appendix C: In-phase locking of a random array of lasers}
\label{sec:AppC}
Figure~\ref{fig:6_Kagome_phs_lck} of the manuscript showed in-phase locking of lasers in a Kagome array. In this section, we show in-phase locking in a random array of lasers. A random array of lasers is defined as an array of lasers where each laser is randomly positioned in space with fixed distance $a=300$ $\mu$m between nearest neighbors. Figure~\ref{fig:S4_rand_array} shows the near-field and far-field intensity distributions of in-phase locked lasers in a random array, when using a Gaussian aperture. As evident from the central sharp spot in the far-field intensity distribution, lasers are in-phase locked with a relatively long coupling range.

\begin{figure}[!ht]
\centering\includegraphics[width=0.73\textwidth]{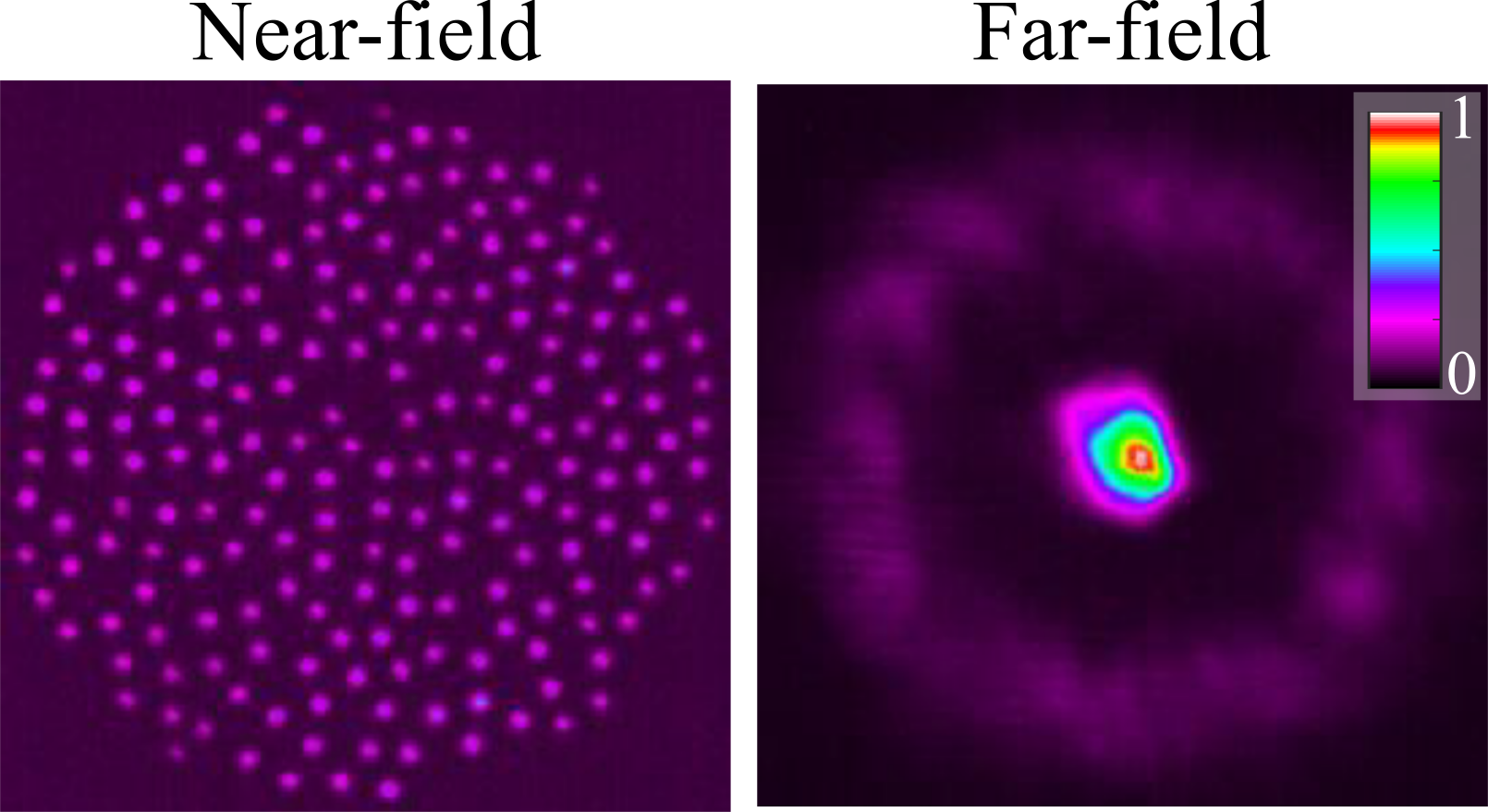}
\caption{\textbf{In-phase locking in a random array of lasers with a Gaussian aperture.} Detected near-field and far-field intensity distributions.}
\label{fig:S4_rand_array}
\end{figure}

\section{Appendix D: Fabrication of Gaussian aperture}
\label{sec:AppD}
\label{sec:manufacturing_G}
In this section, we describe the fabrication of the Gaussian aperture. Figure~\ref{fig:S5_Gaussian_ap_fabric} shows microscope images of two Gaussian apertures of diameter $D=0.5$ mm (Fig.~\ref{fig:S5_Gaussian_ap_fabric}a) and $D=1.5$ mm (Fig.~\ref{fig:S5_Gaussian_ap_fabric}b). The Gaussian apertures were fabricated by HTA Photomask corporation. A total of $16$ Gaussian apertures were designed with diameters ranging from $[0.3$ to $1.8]$ mm, all on a fused silica plate of $4"\times4"\times0.06"$. Each Gaussian aperture is composed of a many discrete steps of $dr=5\pm0.5$ $\mu$m resolution and $256$ grayscale levels, where level $0$ corresponds to transmission of $0$ and level $256$ corresponds to transmission of $1$.

The transmission function of a Gaussian aperture was implemented by depositing a layer of chrome of $100$ nm thickness on the fused silica substrate plate and then using laser writing technique. When coated with chrome, the plate has a transmission of $0.2\%$ for ultra-violet and visible lights. The laser writing technique removes $5\pm0.5$ $\mu$m squares of chrome from the substrate. To implement transmission ranging from $0$ to $1$, the density of chrome squares is varied, as shown in Fig.~\ref{fig:S5_Gaussian_ap_fabric}. To implement a transmission of $1$, as in the center of the Gaussian aperture, all the chrome squares are removed. Then gradually (with $256$ grayscale levels), the density of squares of chrome is increased to implement lower transmissions. To implement a transmission of $0.5$, as at the half maximum of the Gaussian aperture, the density of chrome squares is halved. Finally, to implement a transmission of about $0$, as outside the Gaussian aperture, no chrome squares are removed.

\begin{figure}[!ht]
\centering\includegraphics[width=0.8\textwidth]{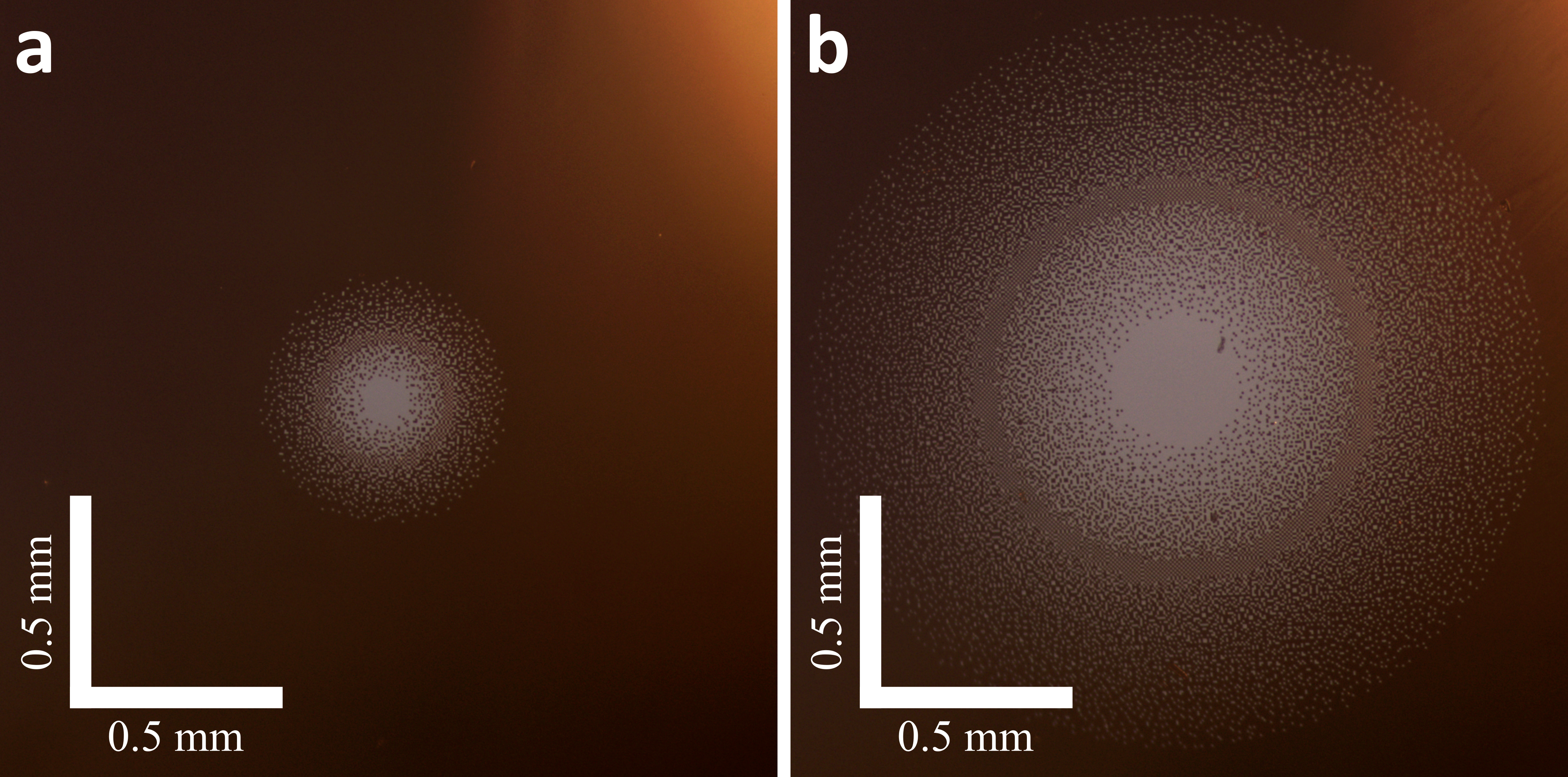}
\caption{\textbf{Microscope images of two Gaussian apertures.} \textbf{a} $D=0.5$ mm. \textbf{b} $D=1.5$ mm.}
\label{fig:S5_Gaussian_ap_fabric}
\end{figure}

\section{Appendix E: Power efficiency of Gaussian coupling}
\label{sec:AppE}
To determine the power efficiency with different couplings, we performed numerical simulations of a DCL without an aperture, with a binary aperture and with a Gaussian aperture, for different aperture diameters. The results, in Fig.~\ref{fig:S6_Power_Efficiency}, show the output power as a function of the aperture diameter for no aperture (black dashed line), a binary aperture (blue line with plus signs), and a Gaussian aperture (orange line with cross signs). Then, the output powers were normalized by the output powers of a DCL with no aperture.

\begin{figure}[!ht]
\centering\includegraphics[width=0.9\textwidth]{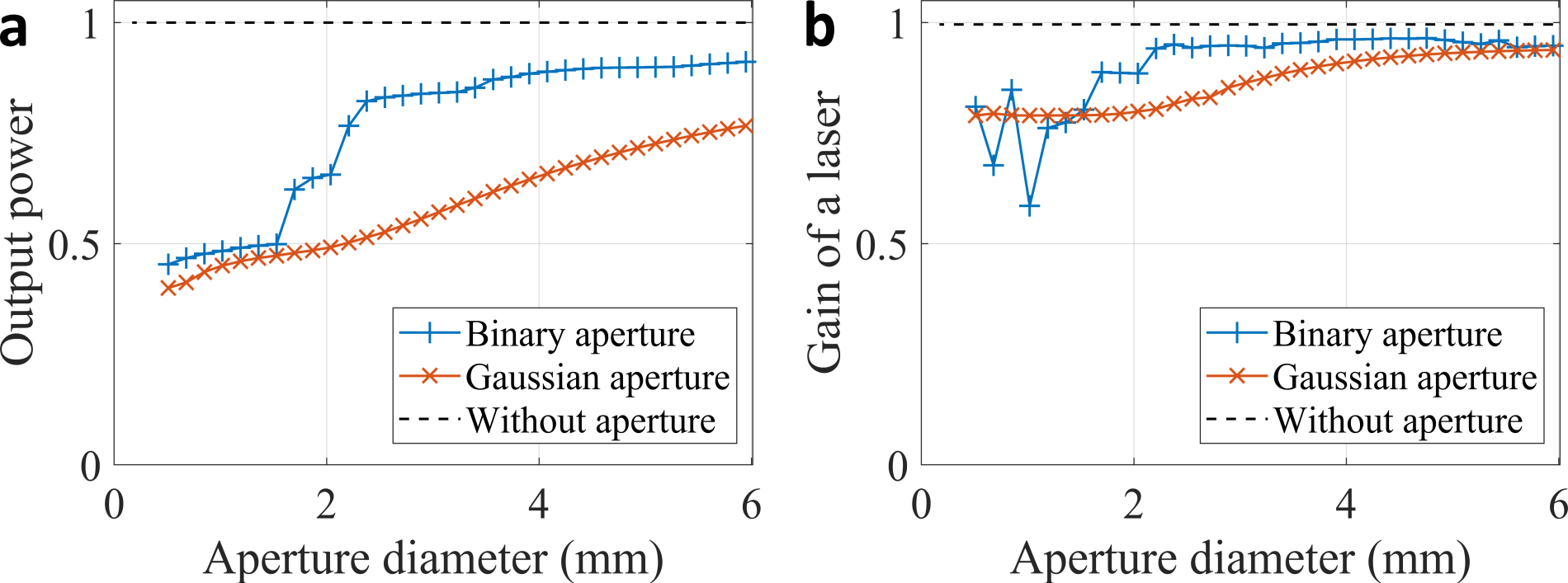}
\caption{\textbf{Simulated power efficiency of the degenerate cavity laser with binary apertures or Gaussian apertures of different sizes.} \textbf{a} Normalized output power. \textbf{b} Normalized gain of a laser in the array.} 
\label{fig:S6_Power_Efficiency}
\end{figure}

As evident in Fig.~\ref{fig:S6_Power_Efficiency}a, the output power with a binary aperture or with a Gaussian aperture is always lower than with no aperture. For large $D=6$ mm aperture diameter, the output power with a binary aperture was about $90\%$, somewhat higher than the $80\%$ with a Gaussian aperture. This difference in output power is due to the fact that the Gaussian aperture reduces internal phase structures of lasers (Fig.~\ref{fig:S8_no_internal_structure}), causing additional loss.

As the aperture diameter decreased from $D=6$ mm to $D=3.5$ mm, the output power with a binary aperture marginally decreased by few percent, while the output power with a Gaussian aperture monotonically decreased by almost $20\%$. This is because from $D=6$ mm to $D=3.5$ mm, with a binary aperture, the lasers remained uncoupled. While, with a Gaussian aperture, the lasers started to couple positively with a continuous increase in the range of the coupling, and at $D=3$ mm, the lasers were already coupled with relatively long range, as shown in Fig.~\ref{fig:4_B_vs_G_coupling}a.

At $D=3.5$ mm, a decrease of $3\%$ occurred in the output power with a binary aperture, corresponding to the transition from uncoupled lasers to negatively coupled lasers. From $D=3.5$ mm to $D=2.4$ mm, the output power with a binary and with a Gaussian apertures monotonically decreased.

At $D=2.3$ mm, the output power with a binary aperture experienced a second sharp decrease of almost $20\%$, corresponding to the transition from negatively coupled (frustrated) lasers to positively coupled lasers, as shown in Figs.~4b and 4c. The range of the coupling at $D=2$ mm was short with an output power of $65\%$, similar to the range of positively coupled laser with a Gaussian aperture of diameter $D=4$ mm (Fig.~\ref{fig:5_pro_G_coupling}a) with a similar output power of $65\%$. Thereby, for equivalent short range in-phase locking of lasers, the output powers with a binary and with a Gaussian apertures are equivalent. Both phase-locking methods thereby experienced the same power efficiency numerically.

From $D=2.3$ mm to $D=1.5$ mm, the output power with a binary aperture sharply decreased while the output power with a Gaussian aperture monotonically decreased. During this transition ($D=2.3$ mm to $D=1.5$ mm), the range of the coupling with a binary aperture increased from short to long coupling range, as shown in Fig.~\ref{fig:4_B_vs_G_coupling}d. At $D=1.5$ mm, the output powers with a binary and with a Gaussian were equivalent, about $50\%$. At $D=1$ mm, the ranges of the positive coupling were also equivalent, as shown in Fig.~\ref{fig:4_B_vs_G_coupling}d.

Finally, from $D=1.5$ mm to $D=0.5$ mm, the output powers with a binary aperture and with a Gaussian aperture have a similar monotonic decrease.

Figure~\ref{fig:S6_Power_Efficiency}b shows the gain of a selected laser at the center of the array as a function of the aperture diameter for no aperture (black dashed line), binary aperture (blue line with plus signs), and Gaussian aperture (orange line with cross signs). The gains are normalized by the gain of a DCL with no aperture. As evident, a similar behavior to that with the output power in Fig.~\ref{fig:S6_Power_Efficiency}a is observed, where the binary aperture gain experiences two sharp decreases at $D=3.5$ mm and $D=2.3$ mm while the Gaussian aperture gain monotonically decreases. The fluctuations at small binary apertures ($D < 1.5$ mm) are because the aperture sizes are becoming too small for the lasers to support lasing.

Next, we investigated the extra loss induced by the first-order sharp spots in the far-field intensity distribution of in-phase locked lasers. In the far-field intensity distribution of Fig.~\ref{fig:4_B_vs_G_coupling}d, the four spots, located at distance $L_{1,0}=\lambda f/a=1.4$ mm away from the central zero-order sharp spot, correspond to first-order sharp spots. These spots contain less power than the zero-order central sharp spot, but they still contain significant power. Since they are outside the range of the Gaussian aperture, they are blocked by the aperture, causing extra loss. This issue can be easily resolved by designing a first-order Gaussian aperture.

A first-order Gaussian aperture combines a zero-order Gaussian aperture with four other  Gaussian apertures but each of them that are shifted by a distance of $L_{1,0}$ in one of the four directions (left, right, up or down), see inset in Fig.~\ref{fig:S7_Power_Efficiency_2}a. To avoid transmissions greater than $1$, an intensity cutoff is applied to the first-order Gaussian aperture. Figure~\ref{fig:S7_Power_Efficiency_2} shows the power efficiency results for no aperture, first-order binary apertures and first-order Gaussian apertures. As evident, the results are similar to those in Fig.~\ref{fig:S6_Power_Efficiency} except that the output power at small aperture diameters is significantly increased to a $75\%$ level. Experimentally, we also measured a significant increase of the output power with a first-order Gaussian aperture over a zero-order Gaussian aperture (not shown).

\begin{figure}[!ht]
\centering\includegraphics[width=0.85\textwidth]{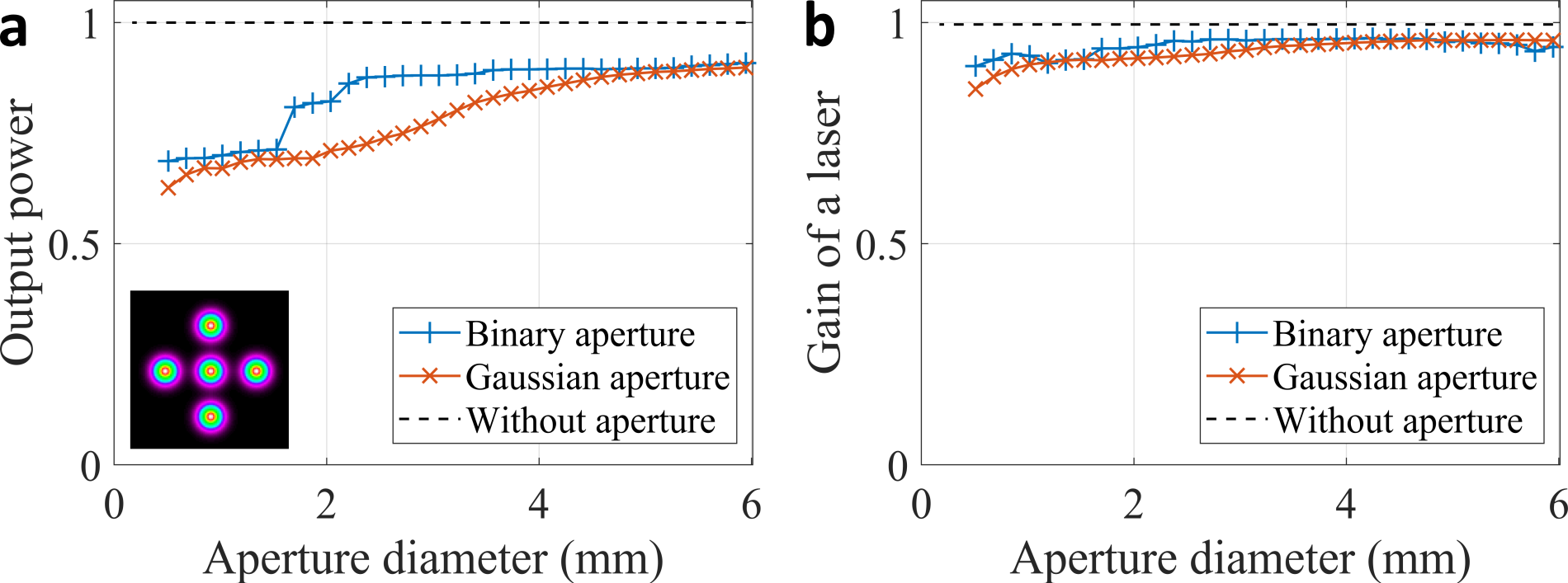}
\caption{\textbf{Simulated power efficiency of the degenerate cavity laser with first-order binary apertures or first-order Gaussian apertures of different sizes.} \textbf{a} Normalized output power. \textbf{b} Normalized gain of a laser in the array. Inset – First-order Gaussian aperture transmission distribution.} 
\label{fig:S7_Power_Efficiency_2}
\end{figure}

\newpage
\section{Appendix F: Generating a pure Gaussian mode with a Gaussian aperture}
\label{sec:AppF}
The far-field Gaussian aperture also acts as a spatial filter, ensuring that each laser in the array is a pure Gaussian mode with no internal structure, as shown in Fig.~\ref{fig:S8_no_internal_structure}. As evident, the insertion of a large Gaussian aperture (so there is no phase-locking) suppressed internal phase structures of each laser and improved the beam quality.

When a laser in an array is not under any phase constraint (as in a standard DCL), it can exhibit different phases (internal phase structures) due to the different modes in the laser, as shown in Fig.~\ref{fig:S8_no_internal_structure}a. Internal phase structures also appear when the lasers are frustrated or weakly coupled. As evident in Fig.~\ref{fig:S8_no_internal_structure}, with no aperture, each laser in the array is composed of many different phases while with a Gaussian aperture, each laser exhibits a single uniform phase. As a consequence, the overall quality of the laser beam is improved with a Gaussian aperture. With no aperture, the intensities of the lasers are composed of a superposition of Gaussian modes. With a Gaussian aperture, the lasers are pure Gaussian modes. 

\begin{figure}[h!]
\centering\includegraphics[width=0.53\textwidth]{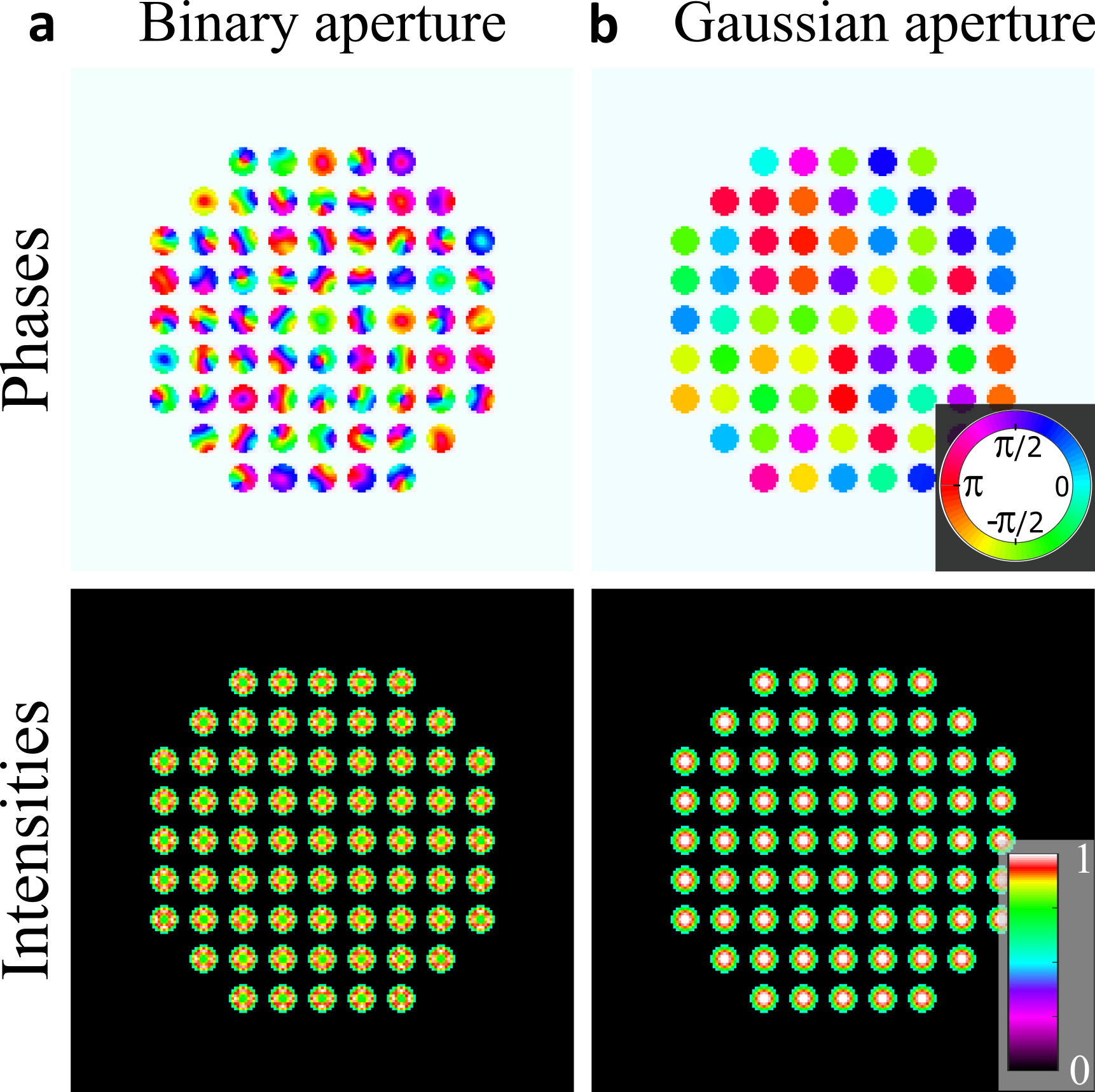}
\caption{\textbf{Simulated phase and intensity distributions of lasers in a square array.} 
\textbf{a} Without an aperture. \textbf{b} With a large Gaussian aperture ($D=6.5$ mm) in the far-field. As evident, the Gaussian aperture suppressed the internal phase structure of each laser and improved the beam quality.}
\label{fig:S8_no_internal_structure}
\end{figure}
\end{backmatter}
\end{document}